\documentclass[12pt, prd, showpacs]{revtex4}
\usepackage{amssymb}
\usepackage{amsmath}

\input{tcilatex}

\begin{document}

\title{High energy collision of particles in the magnetic field far from
black holes}
\author{O. B. Zaslavskii}
\email{zaslav@ukr.net }
\affiliation{Department of Physics and Technology, Kharkov V.N. Karazin National
University, 4 Svoboda Square, Kharkov 61022, Ukraine}
\affiliation{Institute of Mathematics and Mechanics, Kazan Federal University, 18
Kremlyovskaya St., Kazan 420008, Russia}

\begin{abstract}
We consider collision of two particles in the axially symmetric black hole
metric in the magnetic field. If the value of the angular momentum $%
\left\vert L\right\vert $ of one particles grows unbound (but its Killing
energy remains fixed) one can achieve unbound energy in the center of mass
frame $E_{c.m.}$ In the absence of the magnetic field, collision of this
kind is known to happen in the ergoregion. However, if the magnetic field
strength $B$ is also large, with the ratio $\left\vert L\right\vert /B$
being finite, large $E_{c.m.}$ can be achieved even far from a black hole,
in the almost flat region. Such an effect also occurs in the metric of a
rotating star.
\end{abstract}

\maketitle

\section{Introduction}

High-energy collisions near black holes attract much interest both from
theoretical and astrophysical viewpoints. This issue was started in \cite%
{pir1} - \cite{pir3} and revived after an important observation made in \cite%
{ban}. It turned out that the energy $E_{c.m.}$ in the center of mass frame
can grow unbound when two particles moving towards a black hole collide near
the horizon. In doing so, a black hole has to be rotating for the effect to
take place. This is so-called the Ba\~{n}ados - Silk \ - West effect (the
BSW effect, after the names of its authors). Meanwhile, there are
alternative mechanisms connected with the presence of the electromagnetic
field. For the electrically charge nonrotating black hole the similar effect
was obtained in \cite{jl}. The counterpart of the BSW effect due to the
magnetic field in the Schwarzschild background was proposed in \cite{fr}
that gave rise to several works \cite{weak} - \cite{isco} in which more
general metrics were considered. All these works on collisions in the
magnetic field share the common \ feature: for getting large $E_{c.m.}$,
collisions should occur near the horizon similarly to the BSW effect.

From the other hand, there exists one more scenario of getting $E_{c.m.}$
proposed in the paper \cite{ergo} (generalized in \cite{mergo}). It consists
in such collisions when one particle carries a large negative angular
momentum. It turned out that such a scenario can be realized away from the
horizon but inside the ergosphere (or on its boundary) only. The goal of the
present work is to suggest one more mechanism that is in a sense some hybrid
of previous ones. It implies that one of particles has a large negative
angular momentum $\left\vert L\right\vert $ and collisions occurs in the
magnetic field. It turns out that if both $\left\vert L\right\vert $ and the
magnetic field strength are large but their ratio is finite, large $E_{c.m.}$
can be achieved outside the ergoregion. Moreover, under some restrictions,
collisions may happen even in the asymptotically flat region, where
deviation of the black hole metric from the Minkowski one are small.
Actually, this means that effect can be obtained even in the star-like
background when the horizon is absent and is far from the threshold of its
formation. This simplifies greatly the conditions of observation since there
are no objections raised in \cite{mc} (see also discussion in \cite{com})
because of proximity to the horizon and concerning the redshift that "eats"
the gain from large $E_{c.m.}$ when debris of collisions reach a remote
near-flat region.

Throughout the paper, we put fundamental constants $G=c=1,$ except some
cases when we write down $G$ explicitly.

\section{Metric and equations of motion}

Let us consider the metric of the form 
\begin{equation}
ds^{2}=-N^{2}dt^{2}+\frac{dr^{2}}{A}+R^{2}(d\phi -\omega dt)^{2}+g_{\theta
}d\theta ^{2}\text{,}  \label{met}
\end{equation}%
where the coefficients do not depend on $t$ and $\phi $. The horizon
corresponds to $N=0$. We also assume that there is an electromagnetic field
with the four-vector $A^{\mu }$ where the only nonvanishing component is
equal to%
\begin{equation}
A^{\phi }=\frac{B}{2}\text{.}  \label{ab}
\end{equation}

In vacuum, this is an exact solution on the Kerr background with $B=const$ 
\cite{wald}, \cite{ag}. However, we discuss a more general situation and
only afterwards substitute the quantities for the Kerr metric.

Let us consider motion of test particles in this background. The kinematic
momentum $p^{\mu }=mu^{\mu }$, where $m$ is the particle's mass,
four-velocity $u^{\mu }=\frac{dx^{\mu }}{d\tau }$, where $\tau $ is the
proper time, $x^{\mu }$ are coordinates. Then, the kinematic momentum $%
p_{\mu }$ and the generalized one $P_{\mu }$ are related according to

\begin{equation}
p_{\mu }=P_{\mu }-qA_{\mu }\text{,}
\end{equation}%
$q$ is the particle's electric charge. Due to the symmetry of the metric, $%
P_{0}=-E$ and $P_{\phi }=L$ are conserved, where $E$ is the energy, $L$ is
the angular momentum. 

We assume that the equatorial plane is a symmetry one and consider motion
constrained within this plane, so $\theta =\frac{\pi }{2}$. Redefining the
new radial coordinate, we can always achieve that 
\begin{equation}
A=N^{2}  \label{an}
\end{equation}%
within this plane. Then, equations of motion give us 
\begin{equation}
\dot{t}=\frac{X}{N^{2}m}\text{,}
\end{equation}%
\begin{equation}
\dot{\phi}=\frac{\beta }{R}+\frac{\omega X}{mN^{2}}\text{,}  \label{ft}
\end{equation}%
\begin{equation}
X=E-\omega L\text{,}  \label{X}
\end{equation}%
\begin{equation}
\beta =\frac{L-L_{0}}{mR}\text{,}  \label{b}
\end{equation}%
\begin{equation}
L_{0}=\frac{qBR^{2}}{2}\text{,}  \label{l0}
\end{equation}%
\begin{equation}
m^{2}\dot{r}^{2}=Z^{2}\text{,}  \label{rt}
\end{equation}%
\begin{equation}
Z^{2}=X^{2}-m^{2}N^{2}(1+\beta ^{2})\text{.}  \label{v}
\end{equation}%
Dot denotes differentiation with respect to the proper time $\tau $. As
usual, we assume the forward in time condition $\dot{t}>0$, so 
\begin{equation}
X\geq 0.  \label{xt}
\end{equation}

\section{Particle collisions: general formulas}

Let two particles collide. We label their characteristics by indices 1 and
2. Then, in the point of collision, one can define the energy in the centre
of mass frame as%
\begin{equation}
E_{c.m.}^{2}=-p_{\mu }p^{\mu }=m_{1}^{2}+m_{2}^{2}+2m_{1}m_{2}\gamma \text{.}
\end{equation}

Here,%
\begin{equation}
p^{\mu }=m_{1}u_{1}^{\mu }+m_{2}u_{2}^{\mu }
\end{equation}%
is the total momentum,%
\begin{equation}
\gamma =-u_{1\mu }u_{2}^{\mu }
\end{equation}%
is the Lorentz factor of their relative motion.

For motion in the equatorial plane in the external magnetic field (\ref{ab}%
), one finds from the equations of motion (\ref{ft}), (\ref{rt}) that%
\begin{equation}
\gamma =\frac{X_{1}X_{2}-\varepsilon _{1}\varepsilon _{2}Z_{1}Z_{2}}{%
m_{1}m_{2}N^{2}}-\beta _{1}\beta _{2}\text{.}  \label{ga}
\end{equation}

Here, $\varepsilon =+1$, if the particle moves away from the horizon and $%
\varepsilon =-1$, if it moves towards it.

To get large $\gamma $ and $E_{c.m.}$, there are two ways - either to
decrease the denominator or to increase the numerator. The first way implies
that collision should occur near the horizon, where $N\ll 1$. The second one
corresponds to collisions with large $\left\vert L\right\vert $, when the
numerator becomes large as well. In the absence of the magnetic field, such
a scenario can be realized inside the ergosphere or on its boundary only 
\cite{ergo}, \cite{mergo}. Now, we will show that the magnetic field strong
enough extends this scenario outside the ergosphere.

To make the issue nontrivial, we are interested in the case when large $%
E_{c.m.}$ can be obtained with finite individual energies $E_{1}$ and $E_{2}$
only.

\section{Large B and L}

Let us try to achieve large $\gamma $ at the expense of large $L$. We assume
that $\omega >0$, like for the Kerr metric. Then, it follows from (\ref{X}),
(\ref{xt}) that if the angular momentum is large it should be negative,

\begin{equation}
L=-\left\vert L\right\vert \text{.}
\end{equation}%
Let formally $\left\vert L\right\vert \rightarrow \infty $. This means that
in (\ref{X}), 
\begin{equation}
\omega \left\vert L\right\vert \gg E\text{.}  \label{lx}
\end{equation}

Then, the condition $Z^{2}>0$ in (\ref{v}) gives rise to the inequality \ 
\begin{equation}
\frac{\omega R}{N}\left\vert L\right\vert >\left\vert \left\vert
L\right\vert +L_{0}\right\vert .  \label{neq}
\end{equation}

If $L_{0}>0$, this entails 
\begin{equation}
\left\vert L\right\vert (\frac{\omega R}{N}-1)>\left\vert L_{0}\right\vert 
\text{,}
\end{equation}%
whence%
\begin{equation}
\omega R>N\text{.}  \label{er}
\end{equation}%
According to (\ref{met}),%
\begin{equation}
g_{00}=-N^{2}+\omega ^{2}R^{2}\text{.}
\end{equation}%
$.$

Thus we have from (\ref{er}) that $g_{00}>0$. In other words, this is just
the ergoregion, so we return to the situation considered in \cite{ergo}, 
\cite{mergo}.

Let now $L_{0}<0.$

a) $\left\vert L\right\vert >\left\vert L_{0}\right\vert $

If we write 
\begin{equation}
\left\vert L\right\vert =\xi \left\vert L_{0}\right\vert ,  \label{ksi}
\end{equation}%
we obtain from (\ref{neq}) that%
\begin{equation}
\xi >1\text{, }\xi (1-\frac{\omega R}{N})<1\text{.}  \label{a}
\end{equation}

b) $\left\vert L\right\vert \leq \left\vert L_{0}\right\vert $

In a similar way,%
\begin{equation}
\frac{1}{1+\frac{\omega R}{N}}<\xi \leq 1\text{.}  \label{br}
\end{equation}

In both cases a) and b), the sign of the quantity $N-\omega R$ can be
arbitrary, so motion is possible inside or outside the ergoregion.

If the metric is asymptotically flat, $N\rightarrow 1$ and $\omega
\rightarrow 0$ when $R\rightarrow \infty $. We assume that, moreover, $%
\omega R\rightarrow 0$ (like in the case of the Kerr metric - see below).
Then, it is seen from (\ref{a}), (\ref{br}) that motion with large $L$ (\ref%
{ksi}) in this region is possible if $\xi \approx 1$ only. This means that
the angular momentum is fine-tuned according to (\ref{l0}). In this sense,
the value $\xi =1$ is the analogue of the critical trajectory in the BSW
effect \cite{ban}, \cite{prd}. However, for finite $r$, any $\xi $ obeying (%
\ref{a}) or (\ref{br}) is suitable.

\section{Behavior of Lorentz factor}

Let particle 1 have an angular momentum (\ref{ksi}) with $\left\vert
L_{0}\right\vert \rightarrow \infty $. Then, we obtain from (\ref{ga})%
\begin{equation}
\gamma \approx \left\vert L_{0}\right\vert [\frac{X_{2}\omega \xi
-\varepsilon _{1}\varepsilon _{2}Z_{2}\sqrt{\omega ^{2}\xi ^{2}-\frac{N^{2}}{%
R^{2}}(1-\xi )^{2}}}{m_{1}m_{2}N^{2}}-\frac{\beta _{2}(1-\xi )}{Rm}]\text{.}
\end{equation}

As we will be interested in large $r$, we put $\xi =1$. Let us also take for
simplicity $q_{2}=0$, $m_{1}=m_{2}=m=E_{2}\,$, $L_{2}=0$. Then, 
\begin{equation}
\gamma \approx \frac{\left\vert L_{0}\right\vert }{m^{2}N^{2}}\omega
(E_{2}-\varepsilon _{1}\varepsilon _{2}\sqrt{E_{2}^{2}-m^{2}N^{2}})
\end{equation}

Assuming that $E_{2}\sim m$ and omitting the factor of the order unity, we
can write%
\begin{equation}
\gamma \sim \frac{\left\vert L_{0}\right\vert \omega }{mN^{2}}\text{.}
\label{gam}
\end{equation}

\section{Kerr metric}

Now, we apply the above results to the Kerr metric using eq. (\ref{gam}).

In terms of the Boyer-Lindquiste coordinate $r$, the coefficients of the
Kerr metric in the form (\ref{met}) read%
\begin{equation}
\frac{\omega }{N^{2}}=\frac{2Ma}{r(r^{2}-2Mr+a^{2})}\text{,}
\end{equation}%
\begin{equation}
R^{2}=r^{2}+a^{2}+\frac{2M}{r}a^{2}\text{,}
\end{equation}%
where $M$ is the black hole mass, $a=\frac{J}{M}$, $J$ is its angular
momentum. One should take into account that in the point of observation, $%
L_{0}$ is given by (\ref{l0}). Considering large $r\gg 2M$, we obtain%
\begin{equation}
\gamma \approx \frac{ab}{r}\text{,}
\end{equation}%
\begin{equation}
b\equiv \frac{\left\vert qB\right\vert GM}{m}\text{,}  \label{bq}
\end{equation}%
where we restored the gravitational constant $G$.

To achieve $\gamma \gg 1$, we must have%
\begin{equation}
2GM\ll r\ll Gab\text{,}  \label{abr}
\end{equation}%
so 
\begin{equation}
b\gg 1\text{.}  \label{bg}
\end{equation}

One can check that, under the condition (\ref{abr}), the condition (\ref{lx}%
) is satisfied automatically.

If $b$ is large enough, the second inequality in (\ref{abr}) is not
restrictive. To achieve large $\gamma $, particle 1 should have $\left\vert
L\right\vert \sim \left\vert L_{0}\right\vert $. In terms of $b$, 
\begin{equation}
L_{0}=bm\frac{R^{2}}{2M}\text{.}
\end{equation}%
For $R\gtrsim 2M$, $L_{0}\gtrsim 2mMb$. Thus its value must increase by the
factor $b$ as compared to the value of the order $2mM$ typical of motion of
test particles within the Kerr background. Depending on $M$ and $B$, one can
choose $b$ big enough to get $\gamma \gg 1$ but, at the same time, make $%
\frac{L_{0}}{2mM}$ not tremendous. Now, we have an additional parameter $B$
that is absent in the case of collisions without the magnetic field. Varying 
$B$, one can hope to facilitate realization of the scenario under discussion.

\section{Discussion amd conclusion}

In the most part of works on the BSW effect or its modification, unbound $%
E_{c.m.}$ were obtained due to the proximity of collision to the event
horizon or (if the horizon is absent) due to proximity of the would-be
horizon to the threshold of its formation. A step away from this requirement
was made in \cite{ergo}, where collision could take place not in the
vicinity of the horizon. However, this implied the region of strong
gravitational field since the corresponding mechanism works in the
ergoregion (or, in the ultimate case, on its boundary). In this sense, the
region of strong gravitational field was still needed. In the present work,
we made, in a sense, one more step away from the horizon and showed that
even the ergosphere is not necessary. Moreover, collision can occur in the
region, where gravitational field is weak and, nonetheless, we can achieve $%
\gamma \gg 1$. From the other hand, we cannot simply take the limit $%
G\rightarrow 0$ since the conditions (\ref{abr}), (\ref{bg}) should be
satisfied and, additionally, $G$ enters the definition of $b$ (\ref{bq}). In
other words, the gravitational field causes small (but essentially nonzero)
perturbation of the Minkowski metric but affects motion of test particles
significantly. In a similar way, the magnetic field is also weak (since it
almost does not change the metric) and strong (since it is important for
test particles) simultaneously in different aspects \cite{fr}.

It is worth also noting that in the previous scenarios of high energy
collisions in the magnetic and gravitational fields \cite{fr}, \cite{weak}
either collision occurred near the innermost stable circular orbit (ISCO) or
one of particles had the parameters typical of such an orbit.
Correspondingly, motion on ISCO posed restriction on the relationship
between $E$ and $L$. In our case, we considered collision with arbitrary $E$
and $L$ which are independent quantities. The energy $E$ is finite but $L$
should be negative and large enough with $\left\vert L/B\right\vert $ being
finite. (The situation when $E$ is finite but $L$ grows with $B$ is typical
also for collisions near the magnetized Schwarzschild black hole \cite{fr}
and, in some cases, also for the Kerr black hole \cite{weak}.)

The fact that large $\gamma $ can be achieved far from a black hole,
actually means that instead of a black hole we can take a rotating star, so
the background may look unlike a black hole but, nonetheless, produce the
effect under discussion.

That collision can occur far from the black hole, gives hope that
high-energy collisions in gravitational and magnetic fields (each of which
is strong in one sense and weak in another one) can represent not only
theoretical interest but can be relevant for relativistic astrophysics.


\begin{thebibliography}{99}
\bibitem{pir1} T. Piran, J. Katz, and J. Shaham, Astrophys. J. \textbf{196},
L107 (1975).

\bibitem{pir2} T. Piran and J. Shaham, Astrophys. J. \textbf{214}, 268
(1977).

\bibitem{pir3} T. Piran and J. Shanam, Phys. Rev. D \textbf{16}, 1615 (1977).

\bibitem{ban} M. Ba\~{n}ados, J. Silk and S.M. West, Phys. Rev. Lett. 
\textbf{103,} 111102 (2009) [arXiv:0909.0169].

\bibitem{jl} O. Zaslavskii, Pis'ma ZhETF \textbf{92}, 635 (2010) (JETP
Letters \textbf{9}2, 571 (2010)), [arXiv:1007.4598].

\bibitem{fr} V. P. Frolov, Phys. Rev. D\textbf{\ 85}, 024020 (2012)
[arXiv:1110.6274].

\bibitem{weak} T. Igata, T. Harada, and M. Kimura, Phys. Rev. D \textbf{85},
104028 (2012) [arXiv:1202.4859].

\bibitem{string} A.A. Tursunov, \ M. Kolo\v{s}, A.A. Abdujabbarov, B.J.
Ahmedov, and Z. Stuchl\'{\i}k, Phys. Rev. D \textbf{88}, 124001 (2013)
[arXiv:1311.1751].

\bibitem{mpl} O. B. Zaslavskii, [arXiv:1403.6286]. To appear in MOd. Phys.\
Lett. A.

\bibitem{isco} O. B. Zaslavskii, [arXiv:1405.2543].

\bibitem{ergo} A. A. Grib and Yu. V. Pavlov, Europhys. Lett. 101, 20004
(2013) [arXiv:1301.0698].

\bibitem{mergo} O. B. Zaslavskii, Mod. Phys. Lett. A. Vol. 28, No. 11 (2013)
1350037 [arXiv:1301.4699].

\bibitem{mc} S. T. McWilliams, Phys. Rev. Lett. \textbf{110}, 011102 (2013)
[arXiv:1212.1235].

\bibitem{com} O. B. Zaslavskii, Phys. Rev. Lett. \textbf{111}, 079001 (2013)
[arXiv:1301.3429].

\bibitem{wald} R. M. Wald, Phys. Rev. D \textbf{10}, 1680 (1974).

\bibitem{ag} A. N. Aliev and D.V. Gal'tsov, Sov. Phys. Usp. \textbf{32}, 75
(1989).

\bibitem{prd} O.B. Zaslavskii, Phys. Rev. \textbf{D 82} (2010) 083004
[arXiv:1007.3678].
\end{thebibliography}
\end{document}